\documentclass[aps,prd,twocolumn,groupedaddress,showpacs]{revtex4}
\usepackage{graphicx}
\usepackage{dcolumn}
\usepackage{bm}
\usepackage{amssymb,amsmath,latexsym,amsfonts}

\begin{document}

   \title{Planck--Scale Traces from the  
  Interference Pattern of two Bose--Einstein Condensates}

\author{E. Castellanos \footnote{\textbf{Dedicated to the loving memory of my Father, El\'ias Castellanos de Luna, RIP 2014.}}}
\email{ecastellanos@mctp.mx} \affiliation {Mesoamerican Centre for Theoretical Physics \\ (ICTP regional headquarters in Central America, the Caribbean and Mexico) \\ Universidad Aut\'onoma de Chiapas.\\
Ciudad Universitaria, Carretera Zapata Km. 4, Real del Bosque (Ter\'an), 29040, \\Tuxtla Guti\'errez, Chiapas, M\'exico.}

\author{ J. I. Rivas}
\email{jirs@xanum.uam.mx} \affiliation{Departamento de F\'isica, Universidad Aut\'onoma
Metropolitana-Iztapalapa,\\
A. P.  55-534, 09340 M\'exico D.
F., M\'exico.}

\begin{abstract}
We analyze the possible effects arising from Planck scale regime upon the interference pattern of two non--interacting Bose--Einstein condensates. We start with the analysis of the free expansion of a condensate, taken into account the effects produced by a deformed dispersion relation, suggested in several quantum--gravity models. The analysis of the condensate free expansion, in particular, the \emph{modified} free velocity expansion, suggests in a natural way, a modified uncertainty principle that could leads to \emph{new} phenomenological implications related to the quantum structure of space time. Finally, we analyze the corresponding separation between the interference fringes after the two condensates overlap, in order to explore the sensitivity of the system to possible signals caused by the Planck scale regime.
\end{abstract}

\pacs{04.60.Bc, 04.90.+e, 05.30.Jp}
\maketitle

\section{Introduction}

Recently, the use of many--body systems as theoretical tools in searching some possible Planck scale manifestations has become a very interesting line of research \cite{echa,Eli1,echa1,I,JD,AM}. In particular, due to its quantum properties, and also to its high experimental precision, Bose--Einstein condensates become an excellent tool in the search of traces from Planck--scale physics, and has produced several interesting works in this direction \cite{echa,Eli1,echa1,CastellanosClaus,Castellanos,r1,r2,cam1}, and references therein.

First of all, in Refs. \cite{I,JD}, for instance, it was argued that a modified uncertainty principle, could be used to explore some properties of the center of mass motion of macroscopic bodies, which could lead to observable manifestations of Planck scale physics in low energy earth--based--experiments. However, in Ref.\,\cite{AM}, it was suggested that the extrapolation of Planck scale quantization to macroscopic bodies is \emph{incorrect}, due to the fact that these possible manifestations, would be more weakly for macroscopic bodies than for its constituents. This last conclusion comes from the fact that the corrections caused by the quantum structure of space--time, on the properties associated with the center of mass motion of the macroscopic body, seems to be suppressed by the number of particles $(N)$, composing the system. In other words, as it was argued in Ref.\,\cite{AM}, this simple analysis suggests that the possible signals arising from Planck scale quantization, are more weakly for macroscopic bodies than for its own constituents.

Nevertheless, the argument exposed in Ref. \cite{AM}, seems to be not a generic criterion, at least for some properties associated with Bose--Einstein condensates. For instance, in Refs. \cite{Eli1,echa1} it was demonstrated that the corrections arising from the quantum structure of space--time, characterized by a deformed dispersion relation, on some relevant properties associated with a Bose--Einstein condensate scales as a non--trivial function of the number of particles.

As mentioned above, the use of Bose--Einstein condensates open an alternative scenario in searching some possible Planck scale signals, through a deformed dispersion relation in low--energy earth--based experiments. In fact, the analysis of some relevant properties associated with a homogeneous condensate, \emph{i.e.}, a condensate in a box, for instance, the corresponding ground state energy, and consequently the pressure and the speed of sound \cite{echa}, present corrections caused by the quantum structure of space--time, which scales as a non--trivial function of the number of particles. Additionally, it is quiet remarkable that the inclusion of a trapping potential improves the sensitivity to Planck scale signals, compared to a condensate in a box \cite{echa1}. 
These facts suggest that the properties associated with many--body systems, in particular some properties associated with a Bose--Einstein condensate could be used, in principle, to obtain representative bounds on the deformation parameters \cite{echa,Castellanos,r1,r2} or to explore the sensitivity for these systems to Planck scale signals \cite{Eli1,echa1,CastellanosClaus,Conti}.  Thus, it is quite interesting to explore the sensitivity to Planck scale signals on certain properties of the condensate, in which the corrections caused by the quantum structure of space-time can be amplified, instead of being suppressed. 

On the other hand,  it is generally accepted that the dispersion relation between the energy $\epsilon$ and the modulus of momentum $p$ of microscopic particles, should be modified due to the quantum structure of space--time \cite{ami,Giovanni1,Claus,Claus1}.  Such a deformed dispersion relation in the non--relativistic limit can be generically 
expressed in ordinary units as follows \cite{Claus,Claus1}
\begin{equation}
\epsilon \simeq
mc^2+\frac{p^{2}}{2m}+\frac{1}{2M_{p}}\Bigl(\xi_{1}mcp+\xi_{2}p^{2}+\xi_{3}\frac{p^{3}}{mc}\Bigr),
\label{ddr}
\end{equation}
where $c$ is the speed of light, and $M_{p}$ ($\simeq 2.18 \times
10^{-8} Kg$) is the Planck mass. The three parameters $\xi_{1}$, $\xi_{2}$, and $\xi_{3}$, are model
dependent \cite{Giovanni1,Claus}, and should take\, positive or
negative values close to $1$. There are some evidence within the
formalism of Loop quantum gravity \cite{Claus,Claus1,5,12} that
indicates a non--zero values for the three parameters, $\xi_{1},\,
\xi_{2},\, \xi_{3}$, and particularly \cite{5,13} that produces a
linear--momentum term in the non--relativistic limit. Unfortunately,
as is usual in a possible quantum gravity phenomenology, the possible bounds
associated with the deformation parameters, open a wide range of
possible magnitudes, which is translated to a significant challenge.

Indeed, the most difficult aspect in searching experimental hints relevant for the quantum-gravity problem is the
smallness of the involved effects \cite{Kostelecky,amelino1}. If this kind of deformations are
characterized by some Planck scale, then the quantum gravity effects
become very small for a single particle \cite{Giovanni1,Claus}.
It is precisely in this direction that  some many--body properties associated with Bose--Einstein condensates, could be helpfully to improve the sensitivity of possible effects caused by the quantum structure of space--time.  

Here it is noteworthy to mention that one of the more interesting phenomena related to Bose--Einstein condensates, is the interference pattern when two condensates overlap \cite{Pethick,Andr}. The interference pattern is a manifestation of the wave (quantum) nature of these many--body systems, and could be produced even when the two condensates are initially completely decoupled. Then, after switching off the corresponding traps, this allow the systems expand, overlap, and eventually produce interference fringes. Such an interference pattern was observed in the experiment \cite{Andr}, among others,  where interference fringes with a period of $\sim15 \times10^{-6} $ \emph{meters} were observed after switching off the trapping potential and letting the condensates expand for 40 \emph{milliseconds} and overlap. Indeed, several experiments associated with the interference pattern of condensates in different situations has been made, see for instance \cite{JJ,YC,Mun} and references therein. Let us remark that when the trapping potential is turned off, the free velocity expansion of the cloud corresponds, approximately, to the velocity predicted by the Heisenberg's uncertainty principle \cite{Pethick,Andr}.

 In this aim, we explore the free velocity expansion of the condensate and consequently,  the corresponding interference pattern when two of these systems overlap, assuming that the single particle energy spectrum is given by 
Eq.(\ref{ddr}), taken into account only the leading order deformation, \emph{i.e.,} setting $\xi_{2}=\xi_{3}=0$. Additionally, we are not interested here in the relative phase between the two condensates, which is a non--trivial topic and also deserves deeper analysis. Thus, we restrict ourselves on the analysis of the free expansion of the condensate together with the separation of the interference fringes when two of these systems overlap.

\section{Anomalous dispersion relation and free expansion of the condensate}

In order to explore the properties of the condensate under free expansion, let us propose the following \emph{modified} energy associated with the system
\begin{eqnarray}
\label{EN}
 E(\psi)=\int d \mathbf{r} &\Bigg[&\frac{\hbar^{2}}{2m}|\mathbf{\nabla} \psi(\mathbf{r})|^{2}+V(\mathbf{r})|\psi(\mathbf{r})|^{2} \\\nonumber
 &+&\frac{1}{2}U_{0}|\psi(\mathbf{r})|^{4}+\hbar \alpha |\psi(\mathbf{r})\vert\nabla|\psi(\mathbf{r})\vert\Bigg],
\end{eqnarray}
where $\psi$ is the wave function of the condensate or the so--called order parameter, $V(r)=m\omega_{0}^{2}r^{2}/2$ is the external potential, that we will assume for simplicity as an isotropic harmonic oscillator. The term $U_{0}=\frac{4\pi \hbar^{2}}{m}a$, depicts the interatomic potential, being $a$ the s--wave scattering length \emph{i.e.,} only two-body interactions are taken into account. Notice also that we have introduced the contributions due to the deformation parameter $\alpha=\xi_{1}\frac{mc}{2M_{p}}$, assuming, as mentioned above that $\xi_{2}=\xi_{3}=0$. If we set $\alpha=0$, we recover the usual expression associated with the total energy of the cloud \cite{Pethick}. 

An accurate expression for the total energy of the cloud can be obtained employing, as usual, an \emph{anzats} of the form \cite{Pethick}
\begin{equation}
 \psi(\mathbf{r})=\frac{ N^{1/2}}{{\pi}^{3/4}R^{3/2}} \exp(-r^{2}/2R^{2})\exp(i\phi(r)), \label{TF}
\end{equation}
where $N$ is the corresponding number of particles and $R$ is a characteristic length, that is interpreted as the radius of the system.

Notice that Eq.(\ref{TF}) corresponds to the solution of the \,\, Schr\"odinger equation associated with non--interacting systems,
where the phase $\phi$ can be associated with particle currents \cite{Pethick}. Thus, by inserting the \emph{anzats} (\ref{TF})
in the energy functional (\ref{EN}) we are able to obtain the corresponding energy
\begin{equation}
\label{TE}
E=E_{F}+E_{R},
\end{equation}
where $E_{F}$ is the kinetic energy associated with particle currents
\begin{equation}
\label{EFL}
E_{F}=\frac{\hbar^{2}}{2m} \int d\,\mathbf{r}|\psi(\mathbf{r})|^{2} (\mathbf{\nabla} \phi)^{2}.
\end{equation}
Additionally, $E_{R}$ can be interpreted as the energy associated with an effective potential, which is equal to the total energy of the condensate when the phase $\phi$ does not vary in space. The term $E_{R}$ contains the contributions of the ground state energy ($E_{0}$), the harmonic oscillator potential ($E_{P}$), and the contributions due to the interactions among the particles within the condensate $(E_{I})$. Notice that we have inserted also the contribution  $E_{\alpha}$ caused by the deformation parameter $\alpha$
\begin{equation}
E_{R}=E_{0}+E_{P}+E_{I}+E_{\alpha},
\end{equation}
where
\begin{equation}
\label{EZP}
E_{0}=\frac{\hbar^{2}}{2m}\int d \mathbf{r} \Bigl(\frac{d |\psi (\mathbf{r})|}{dr}\Bigr)^{2},
\end{equation}
\begin{equation}
E_{P}=\frac{1}{2}m\omega_{0}^{2}\int d \mathbf{r} r^{2} |\psi (\mathbf{r})| ^{2},
\end{equation}
\begin{equation}
E_{I}=\frac{1}{2}U_{0}\int d \mathbf{r} |\psi (\mathbf{r})|^{4},
\end{equation}
\begin{equation}
\label{EA}
E_{\alpha}=\hbar \alpha \int d \mathbf{r} \Bigl(\frac{d |\psi (\mathbf{r})|^{2}}{dr}\Bigr).
\end{equation}

Consequently, $E_{R}$ can be written as follows
\begin{eqnarray}
 E_{R}&=&\frac{3}{4}\frac{{\hbar}^{2}}{m {R}^{2}}N+\frac{3}{4}m{{\omega}_{0}}^{2}{R}^{2}N\\ \nonumber
 &+&\frac{{U}_{0}}{2{(2\pi)}^{3/2} R^{3}}{N}^{2}-\alpha\frac{2 \hbar}{\sqrt{\pi}R}N,
\end{eqnarray}
where we have used the trial function (\ref{TF}) together with Eqs. (\ref{EZP})--(\ref{EA}) in order to obtain the above expression.

The equilibrium radius of the system, let say $R_{0}$, can be obtained by minimizing the total energy (\ref{TE}). Additionally, the contribution of the kinetic energy (\ref{EFL}) is positive definite, and is zero when the phase $\phi$ is constant \cite{Pethick}. 

However, when the radius $R$ differs from its equilibrium condition, after the external potential $V(r)=m\omega_{0}^{2}r^{2}/2$ is turned off at, let say $t=0$, there is a force that change $R$ and produces an expansion of the cloud.
In order to determine an equation for the dynamics of the system, we must deduce the corresponding kinetic energy $E_{F}$ in function of time, through its dependence on the radius $R$.
Changing $R$ from its initial value to a new value $\tilde{R}$ amounts to a uniform dilation of the system, since the new density distribution $|\psi(\mathbf{r})|^{2}=n(\mathbf{r})$ may be obtained from the old one by changing the radial coordinate of each atom by a factor $\tilde{R}/R$, see Ref. \cite{Pethick} for details. Thus, the velocity of a particle can be expressed as follows
\begin{equation}
v(r)=r\frac{\dot{R}}{R}\, ,
\end{equation}
where the dot stands for derivative with respect to time. Consequently, the kinetic energy $(E_{F})$ is given by
\begin{equation}
\label{INT}
E_{F}=\frac{mN}{2R^{2}}\frac{\int d\,\mathbf{r}n(\mathbf{r})\, r^{2}}{\int d\, \mathbf{r} n(\mathbf{r})}\,\dot{R}^{2},
\end{equation}
where the ratio between the integrals is a mean--square radius of the condensate \cite{Pethick}.

Then, it is straightforward to obtain the kinetic energy $E_{F}$ by using the \emph{anzats} Eq.(\ref{TF}), with the result $E_{F}=3\dot{R}^{2}Nm/4$. Moreover, assuming that the energy is conserved at any time, we obtain the following energy conservation condition associated with our system
\begin{eqnarray}
\label{ERES}
 &&\frac{3m \dot{R}^{2}}{4}+\frac{3\hbar^{2}}{4mR^{2}}+\frac{{U}_{0}}{2{(2\pi)}^{3/2} R^{3}}{N}
 -\alpha \frac{2 \hbar}{\sqrt{\pi}R}\\\nonumber&=&\frac{3\hbar^{2}}{4mR_{0}^{2}}+\frac{{U}_{0}}{2{(2\pi)}^{3/2} R_{0}^{3}}{N}
 -\alpha \frac{2 \hbar}{\sqrt{\pi}R_{0}},
\end{eqnarray}
where $R_{0}$ is the radius of the condensate at time $t=0$, which is approximately equal to the oscillator
length $a_{ho}=(\hbar/m\omega_{0})^{1/2}$ in the non--interacting case. $R$ is function of time which corresponds to the radius at time $t$. Eq. (\ref{ERES}) must be solved numerically, even in the case $\alpha=0$. However, if we neglect inter--particle interactions, \emph{i.e.}, setting $U_{0}=0$ then, we are able to obtain an analytical solution for the above equation, with the result
\begin{eqnarray}
\label{SOLR}
 &&\frac{1}{{\beta}^{2}}\sqrt{{\beta}^{2}{R}^{2}+\frac{2\hbar\alpha}{\sqrt{\pi}}R-\frac{3{\hbar}^{2}}{4m}}\\ \nonumber
 &-&\frac{\hbar\alpha}{\sqrt{\pi}{\beta}^{3}} \ln\Bigg[\frac{{\beta}^{2}R+\frac{\hbar\alpha}{\sqrt{\pi}}}{{\beta}^{2}{R}_{0}+
 \frac{\hbar\alpha}{\sqrt{\pi}}}+\Bigg({\Bigl( \frac{ {\beta}^{2}R
 +\frac{\hbar\alpha}{\sqrt{\pi}}}{ {\beta}^{2}{R}_{0}
 +\frac{\hbar\alpha}{\sqrt{\pi}}} \Bigr)}^{2}-1\Bigg)^{1/2} \,\, \Bigg] \\ \nonumber
 &=&\sqrt{\frac{4}{3m}}t,
\end{eqnarray}
where we have defined
\begin{eqnarray}
\label{beta}
 {\beta}^{2}=\frac{3{\hbar}^{2}}{4m{R}_{0}^{2}}-\frac{2\hbar\alpha }{\sqrt{\pi}{R}_{0}}.
\end{eqnarray}

A rough approximation for the \emph{modified} width of the packet which is valid for large expansion times and $\alpha<<1$, renders the following solution
 \begin{eqnarray}
\label{SOLR2}
R_{\alpha}^{2}(t) = R_{0}^{2}+\Bigg[\frac{\hbar^{2}}{m^{2}R_{0}^{2}}-\alpha \frac{8}{3 \sqrt{\pi} } \frac{\hbar}{m R_{0}}\Bigg]t^{2}+...\, ,
\end{eqnarray}
 which also is equivalent when $\alpha<<1$ for $R_{\alpha}>>R_{0}$.
If we set $\alpha=0$ then, we recover the usual solution \cite{Pethick}
\begin{equation}
\label{Usual}
{R}^{2}(t)={R}_{0}^{2}+\Bigl(\frac{\hbar }{m R_{0}}\Bigr)^{2}t^{2}.
\end{equation}

Notice that in the usual case, $\alpha=0$, $v_{0}=\frac{\hbar}{mR_{0}}$ is defined as the velocity expansion of the condensate, corresponding to the velocity predicted by the Heisenberg`s uncertainty principle for a particle confined a distance $R_{0}$ \cite{Pethick}. Thus, in the usual case $\alpha=0$, the width of the cloud at time $t$ can be written in its usual form ${R}^{2}(t)={R}_{0}^{2}+(v_{0}t)^{2}$.

On the other hand, from Eq. (\ref{SOLR2}), we are able to define the \emph{square modified velocity expansion} $(v_{0}^{\alpha})^{2}$ as follows
\begin{equation}
\label{VA}
(v_{0}^{\alpha})^{2}=\frac{\hbar^{2}}{m^{2}R_{0}^{2}}-\alpha \frac{8}{3 \sqrt{\pi} } \frac{\hbar}{m R_{0}},
\end{equation}  
which is well defined, since the deformation parameter $\alpha$ has dimensions of velocity. The modification caused by $\alpha$ is quite small, then the following expansion is justified
\begin{equation}
\label{v0}
(v_{0}^{\alpha})=\frac{\hbar}{mR_{0}}-\frac{4}{3\sqrt{\pi}}\alpha+O(\alpha^{2}).
\end{equation}

Here, let us remark that the presence of the deformation parameter $\alpha$ suggests a modification to the Heisenberg's uncertainty principle, which appears in a natural way, just by looking up to the predicted \emph{modified} velocity $(v_{0}^{\alpha})$.
If we define a \emph{new} deformation parameter $\alpha' =\alpha \frac{4m}{3\sqrt{\pi}}$, together with $R_{0}=x$ then, the resulting \emph{modified} uncertainty principle seems to be
\begin{equation}
 \Delta x \Delta p \ge  \frac{\hbar}{2} - \alpha' x+ O(\alpha^{2}).
\end{equation}

Notice that the leading order modification obtained from the analysis of the free expansion of the condensate, is apparently linear in the position which, as far we know, has been not reported in the literature, see for instance Refs. \cite{KEN,AF,BM} and references therein. If so, this fact would open some new phenomenological implications concerning to the quantum--structure of space time. Additionally, it is clear that the parameters $\xi_{2}$ and $\xi_{3}$, also contribute to the funcional form of the modified uncertainty principle. This scenario is a non--trivial topic and deserves deeper investigation, that we will presented elsewhere.

On the other hand, the quantity $h/m$, can be measured by comparing the de Broglie wavelength and the velocity of a particle (which in fact is the velocity predicted by the Heisenberg uncertainty principle), as demonstrated in Ref. \cite{K} in measurements using neutrons. Indeed, the quantity $h/m$ is also related to the velocity $v_{0}$ by the de Broglie equation
\begin{equation}
\frac{h}{m}=\lambda v_{0},
\end{equation}
where $\lambda$ is the corresponding wavelength. The velocity $v_{0}$ of the neutrons is measured using a very precise time--of--flight method, leading to $h/m=3.956 033 332 (290) 10^{-7}$$m^{2}s^{-1}$ and in consequence a precise determination of the fine--structure constant of order $137.036 010 62(5 03)10^{-8}$ was obtained \cite{K}. Both measurements with a relative uncertainty of a few parts in $10^{-8}$.

These ideas were also extended  in Ref.\cite{contrast}, through measurements of the kinetic energy of an atom recoiling due to absorption of a photon using an interferometric technique called "contrast interferometry", in a sodium Bose--Einstein condensate. The quantity $h/m$ can be extracted from a measurement of the photon recoil frequency ($\omega_{r}$) defined as follows \cite{contrast}
\begin{equation}
\label{reco}
\omega_{r}=\frac{\hbar}{2m}k^{2},
\end{equation}
where $k$ is the wavevector of the photon absorbed by the atom, whose value is accurately accessible \cite{Th}. 
There, a measurement of the photon recoil frequency leads to $\omega_{r}= 2\pi \times 24.9973$$kHz$$(1\pm 6.7\times 10^{-6})$. 

Finally, let us add that the form of the energy dispersion relation (\ref{ddr}), was constrained by using high precision atom--recoil frequency measurements \cite{Claus,Claus1}. In such scenario, bounds for the deformation parameters of order $\xi_{1} \sim -1.8 \pm 2.1$ and $|\xi_{2}|\sim 10^{6}$ were obtained.

However, in order to analyze an alternative procedure compared to those used in Refs.\cite{Claus,Claus1}, \emph{i.e.}, by using the \emph{modified} free expansion velocity of the condensate Eq. (\ref{v0}), we are able to obtain the following \emph{modified} the de Broglie equation associated with our system
\begin{equation}
\frac{2\pi \hbar}{m}= R_{0} \Bigl(v_{0}- \alpha \frac{8}{3 \sqrt{\pi}}\Bigr).
\end{equation}
Consequently, the \emph{modified} photon recoil frequency $\omega_{r}^{(\alpha)}$ can be expressed as follows
\begin{equation}
\omega_{r}^{(\alpha)}=\frac{R_{0}} {4 \pi} \Bigl(v_{0}- \alpha \frac{8}{3 \sqrt{\pi}}\Bigr)k^{2}.
\end{equation}
Where we have assumed that the wave vector $\vec{k}$ of the photon absorbed by an atom is independent of the deformation parameter $\alpha$. 

Therefore, the relative shift $(\omega_{r}^{(\alpha)}-\omega_{r})/\omega_{r}\equiv \Delta \omega_{r}^{(\alpha)}/\omega_{r}$ caused by the deformation parameter $\alpha$ is given by
\begin{equation}
\frac{\Delta \omega_{r}^{(\alpha)}}{\omega_{r}}= \alpha \frac{4 R_{0}m}{\pi^{3/2} \hbar}. 
  \label{Spomega}
\end{equation}
The value $\omega_{r}= 2\pi \times 24.9973$$kHz$$(1\pm 6.7\times 10^{-6})$ obtained in Ref. \cite{contrast} together with Eq.(\ref{Spomega}), allows us to obtain a bound for the deformation parameter $\xi_{1}$, under typical laboratory conditions. In such a case we are able to obtain an upper bound up to $|\xi_{1}|\sim 1$, by using the relative shift Eq.(\ref{Spomega})
through its dependence on the \emph{modified} velocity expansion Eq. (\ref{v0}), which is compatible with the upper bound reported in Refs. \cite{Claus,Claus1}.

\section{Interference Pattern of two condensates and Planck scale signals}
Finally, let us analyze the interference pattern of two overlapping Bose--Einstein condensates, in order to explore some possible Planck--scale signals in such phenomenon.
If there is coherence between two condensates, the state may be described by a single condensate wave function, which has the following form
\begin{equation}
\psi_{1,2}(\mathbf{r},t)=\sqrt{N_{1}}\psi_{1}(\mathbf{r},t)+\sqrt{N_{2}}\psi_{2}(\mathbf{r},t),
\end{equation}
where $N_{1}$ and $N_{2}$ corresponds to the number of particles within each cloud.
After the free expansion, the two condensates overlap and interfere. If the effects of interactions are neglected in the overlaping region, the particle density at any point is given by
\begin{eqnarray}
\label{INT1}
n_{1,2}(\mathbf{r},t)&=&|\psi_{1,2}(\mathbf{r},t)|^{2}=N_{1}|\psi_{1}(\mathbf{r},t)|^{2}+N_{2}|\psi_{2}(\mathbf{r},t)|^{2}\,\,\,\,\,\,\,\,\,\,\,\,\,\\\nonumber&+&
2\sqrt{N_{1}N_{2}}Re[\psi_{1}(\mathbf{r},t)\psi^{*}_{2}(\mathbf{r},t)].
\end{eqnarray}
The third right hand term  of expression (\ref{INT1}) corresponds to an interference pattern \cite{Pethick}, caused by the overlap of the two condensates.
In order to obtain the corrections caused by the deformation parameter $\alpha$, on the properties of the interference pattern of two condensates, let us appeal as usual,  to the following time dependent condensate wave functions \cite{Pethick}

\begin{equation}\label{INT2}
 \psi_{1}(\mathbf{r},t)=\frac{e^{i \phi_{1}}}{(\pi R^{2}_{\alpha}(t))^{3/4}} \exp\Bigg[- \frac{(\mathbf{r}-\mathbf{d}/2)^{2}(1-i\hbar t/mR_{0}^{2})}{2R^{2}_{\alpha}(t)}\Bigg],
\end{equation}

\begin{equation}\label{INT3}
 \psi_{2}(\mathbf{r},t)=\frac{e^{i \phi_{2}}}{(\pi R^{2}_{\alpha}(t))^{3/4}} \exp\Bigg[- \frac{(\mathbf{r}+\mathbf{d}/2)^{2}(1-i\hbar t/mR_{0}^{2})}{2R^{2}_{\alpha}(t)}\Bigg],
\end{equation}
where $\phi_{1}$ and $\phi_{2}$ are the initial phases for each condensate, $R_{0}$ is the initial radius of the cloud, which is approximately equal to the oscillator
length $a_{ho}=(\hbar/m\omega_{0})^{1/2}$. Additionally, $R_{\alpha}(t)$ is the is the width of a packet at time t, given by Eq. (\ref{SOLR2}).
If we set $\alpha=0$ in Eqs. (\ref{INT2}) and (\ref{INT3}) then, we recover the usual expressions \cite{Pethick}.

The interference term in Eq. (\ref{INT1}) thus in given by

\begin{eqnarray}\label{INT4}
 Re[\psi_{1}(\mathbf{r},t)\psi^{*}_{2}(\mathbf{r},t)]&=& \frac{{e}^{-\frac{{r}^{2}}{{R}_{\alpha}^{2}(t)}}{e}^{-\frac{{d}^{2}}{4{R}_{\alpha}^{2}(t)}}}{{[\pi{R}_{\alpha}^{2}(t)]}^{3/2}}
 \\ \nonumber &&\times \cos\Bigl(\frac{\hbar}{m}\frac{\mathbf{r}\cdot\mathbf{d}}{{R}_{0}^{2}{R}_{\alpha}^{2}(t)}t+\phi \Bigr).
\end{eqnarray}

Notice that the phase shift $\phi=\phi_{1}-\phi_{2}$ is measurable, although the individual phases $\phi_{1}$ and $\phi_{2}$ are not \cite{WJ}. Here the pre--factor $exp(-{r}^{2}/{R_{\alpha}}^{2}( t ))$ depends slowly on $r$ but the cosine function can give rise to rapid spatial variations. We can notice also from Eq. (\ref{INT4}) that planes of constant phase are perpendicular to the vector between the centers of the two clouds. The positions of the
 maxima depend on the relative phase of the two condensates, and if we take $\mathbf{d}$ to lie in the z direction, the distance between maxima is given by 
\begin{eqnarray}\label{INT5}
z_{(\alpha)}=2\pi\frac{m{R}_{\alpha}^{2}(t){R}_{0}^{2}}{\hbar t d}.
\end{eqnarray}
If the expansion time is sufficiently large, \emph{i.e.}, the cloud has expanded to a size much greater than $R_{0}$ then, as mentioned before, $R_{\alpha}^{2}(t)$ is given approximately by Eq. (\ref{SOLR2}).
Therefore, the distance between maxima associated with the interference fringes is given by the following expression
 \begin{eqnarray}
 \label{DZA}
  z_{(\alpha)}\approx 2\pi\Bigl(\frac{\hbar}{m d}-\frac{8 \alpha
 {R}_{0}}{3\sqrt{\pi}d}\Bigr)t.
 \end{eqnarray}

When $\alpha=0$, we recover the usual result \cite{Pethick,Andr}. In the usual case, $\alpha=0$, the separation between maxima is typically of order $10^{-6}$\,\emph{meters} \cite{Andr}. From relation (\ref{DZA}), we are able to obtain the sensitivity of our system to Planck scale signals upon the fringes separation. Under typical laboratory conditions, \emph{i.e.,} $\omega_{0} \sim 10$Hz  and a typical  mass of order $m \sim
10^{-26}$ \emph{Kilograms}, $d=40\times10^{-6}$\emph{meters}, together with a free expansion time of order $t=40\times10^{-3}$\emph{seconds}, the correction caused by the deformation parameter $\alpha$ can be inferred up to $|\xi_{1}| \times10^{-11}$\,\emph{meters}, \emph{i.e.}, five orders of magnitude smaller than the typical distance between the maxima reported in Ref. \cite{Andr}, when $|\xi_{1}|\sim 1$. 

The above analysis and the form of Eq. (\ref{DZA}), suggests that small(large) separations between the two condensates, together with large(small) expansion times, could be used to search the small traces arising from the quantum structure of the space--time, upon the interference pattern of two Bose--Einstein condensates.
However, in order to obtain a more accurate description to the possible measurement of the contributions caused by the quantum structure of space time, let us analyze the experimental scenario in this context. If the contributions of the Planck scale physics could eventually be measured, this implies that the usual term in Eq. (\ref{DZA}), must to be known more accurated than the size of the correction associated with the deformation parameter 
$\alpha$. Unfortunately, the corresponding experimental error associated with the interference fringes separation is not reported in the literature, at least, in the literature known by the authors. 
In this aim, let us analyze the error propagation in the measure of the fringes separation, when $\alpha=0$ in order to obtain  the experimental conditions that could allow to detect posible signals arising from the Planck scale regime. In other words, in the most unfavorable case this entails
\begin{equation}
\label{error}
\Delta z_{(\alpha=0)}=\Bigg[\Bigl|\frac{\partial z_{(\alpha=0)}}{\partial m }\Bigr|\Delta m+\Bigl|\frac{\partial z_{(\alpha=0)}}{\partial h }\Bigr|\Delta h+\Bigl|\frac{\partial z_{(\alpha=0)}}{\partial d }\Bigr|\Delta d\Bigg]t,
\end{equation}
where as usual, $\Delta z_{(\alpha=0)}$ depicts the experimental error, and so on. Notice that we have re--absorbed for simplicity, the $2\pi$ factor in the definition of the Planck constant $h$. Additionally, the expansion time $t$ can be interpreted here, without lost of generality, as an evolution parameter.
The above expression leads to the following error associated with the fringes separation $\Delta z_{(\alpha=0)}$ in the usual case $\alpha=0$
\begin{equation}
\label{z_0}
\Delta z_{(\alpha=0)}=z_{(\alpha=0)}\Bigl[\frac{m d\Delta h+h d \Delta m+ hm \Delta d}{m h d}\Bigr]t, 
\end{equation}
where $z_{(\alpha=0)}$ is the usual value when $\alpha=0$. The corresponding uncertainties $\Delta m= 0.17$\,\emph{ppb} in atomic mass units for $^{23}Na$, and $\Delta h=20$\,\emph{ppb} in \textbf{SI} units reported in the experiments \cite{MICH}, and \cite{Steele} respectively, can be used to calculate $\Delta z_{(\alpha=0)}$. Unfortunately, as far we know, there is not uncertainty reported for the corresponding initial separation $d$. In this situation we obtain an error for the fringes separation of order $ \Delta z_{(\alpha=0)}\sim(10^{-18}+ 1.5 \times10^{-2}\Delta d)$, for the usual case $\alpha=0$ with $t=40\times10^{-3}$\,\emph{seconds}.
As mentioned above, the order of magnitude associated with $\alpha$ can be inferred up to $|\xi_{1}| \times10^{-11}$
\emph{meters}. This fact implies, in principle, that uncertainties for the distance $d$ of order $\Delta d \ge 6.67\times10^{-10}$
\emph{meters} for expansion times of order $10^{-3}$\,\emph{seconds}, are necessary (assuming $|\xi_{1}|\sim1$), in order to obtain a possible detection of Planck scale signals under typical laboratory conditions.
However, large expansion times up to 4\,\emph{seconds} can be achieved, for instance, in interference free fall experiments \cite{Mun}. In these circumstances, we obtain $\Delta d \ge 2.5\times 10^{-9}$\,\emph{meters}. In other words, large expansion times implies better precision in knowing the initial separation $d$. Notice also that $\Delta z_{(\alpha=0)}$ and $\Delta d$ are, basically, of the same order of magnitude in the cases described above.

In the same spirit, we are capable to calculate the corresponding experimental error ($\Delta z_{\alpha}$) associated with the deformation parameter $\alpha$, \emph{i.e.,} the second term in Eq. (\ref{DZA}) assuming that $M_{p}$ and $c$ are constants, together with $\xi_{1} \sim -1.8 \pm 2.1$ \cite{Claus,Claus1} and $R_{0}=\sqrt{\frac{\hbar}{m\omega_{0}}}$. We assume also that the uncertainty corresponding to typical frequencies is of order $\omega_{0}=21\pm 4$\,\emph{MHz} in the case of magneto--optical traps \cite{tesis}.
Under these conditions, the corresponding uncertainty can be inferred here up to $\Delta z_{\alpha}=10^{-14}$\emph{meters} for $t=40\times10^{-3}$\,\emph{seconds}, assuming $\Delta d \sim 10^{-10}$\,\emph{meters}. Conversely, we obtain $\Delta z_{\alpha}=10^{-11}$\emph{meters} for $t=4$\,\emph{seconds}, assuming $\Delta d \sim 10^{-9}$\,\emph{meters}.
The corresponding errors $\Delta z_{\alpha=0}$ and $\Delta z_{\alpha}$, can be used as a criterion to discriminate how precise is the eventual measurement of the correction term caused by $\alpha$ with respect to the usual term. In fact the above results indicate that if the corrections caused by $\alpha$ wants to be measured, then better precision in needed, compared with the usual term.

The uncertainties obtained for $\Delta z_{\alpha=0}$, can be also used as a criteria to optimize value of $d$. For instance, assuming that $\Delta z_{(\alpha=0)}\sim{10^{-10}}$\,\emph{meters}, corresponding to expansion times of order of $40\times10^{-3}$\,\emph{seconds}, together with the corrections caused by $\alpha$ in Eq. (\ref{DZA}), this leads to initial separations of order $d\sim 6.8\times10^{-8}$\,\emph{meters}. Conversely, if  $\Delta z_{(\alpha=0)}\sim{10^{-9}}$\,\emph{meters} for $t=4$\,\emph{seconds}, this implies  $d\sim{1.15 \times 10^{-10}}$ \emph{meters}. In other words, according to our results, an optimal value for the initial separation $d$ seems to be between $10^{-8}$ and $10^{-10}$\,\emph{meters}.

Finally, let us analyze the relative shift on the fringes separation caused by the deformation term $\alpha$. The relative shift can be expressed as follows 
\begin{equation}
\label{shift}
\frac{z_{(\alpha)}-z_{(\alpha=0)}}{z{_{(\alpha=0)}}}=-\frac{4}{3\sqrt{\pi}}\xi_{1}\sqrt{\frac{m^{3}c^{2}}{\hbar \omega_{0} M_{p}^{2}}},
\end{equation}
where $z_{(\alpha)}$ is given by expression Eq. (\ref{DZA}) and $z_{(\alpha=0)}$ is the usual result, setting $\alpha=0$.  
We notice that the relative shift is apparently independent of the initial separation $d$. The relative shift Eq. (\ref{shift}) can be inferred under typical laboratory conditions up to $\xi_{1}\times 10^{-6}$ for $\omega_{0}\sim 10^{3}$\emph{Hz}, which is approximately of the same order of magnitude than the usual fringes separation when $\alpha=0$, and apparently impossible to be measured. However, let us mention that such a shift can be improved for small $\omega_{0}$. For instance, if $\omega_{0}\sim 10$\emph{Hz} then, the relative shift is of order $\xi_{1} \times 10^{-4}$, for a typical mass $m$ of order $10^{-26}$\emph{Kilograms}, \emph{i.e.,} two orders of magnitud bigger than the typical fringes separation, which is notable.

\section{Conclusions}

We have analyzed the free expansion of a condensate, and also the properties when two of these systems overlap, assuming as a fundamental fact a deformed dispersion relation. We have proved that the free velocity expansion is corrected as a consequence of a posible quantum structure of space time. Additionally, the predicted \emph{modified velocity expansion}, endows in a natural way a modification in the Heisenberg's uncertainty principle, which in principle, open the possibility to explore some phenomenological consequences in other systems and clearly deserves deeper investigation. 

We have explored possible traces arising from Planck scale physics upon the properties associated with the interference fringes when two condensates overlap, and also we have analyzed the experimental scenario under typical laboratory conditions. 
Here it is importan to mention that the contribution caused by interactions among the constituents of the system are expected to be also bigger, evidently, than the contributions caused by the deformation parameter $\alpha$. However, as was recently reported in experiment Ref. \cite{Mun}, the non--linear evolution of the condensate, occurs at very short times ($<30$\emph{milliseconds}). This fact suggests that possible Planck scale signals, could be measured in principle, for times larger than $30$\emph{milliseconds}, in which the system operate deeper in the linear regime \emph{i.e.,} almost in the non--interacting case. In fact, free fall experiments could account for Planck scale signals in this context, in which expansion times of order 4\,\emph{seconds}, can be achieved \cite{Mun}. Nevertheless, the scenario presented in this report must be extended to more realistic situations, in which the contribution caused by the interactions among the constituents of the condensate could be representative, together with the presence of a gravitational field.

Finally, we must add that the possible detection of these corrections, could be out of the current technology. However, it is remarkable that an adequate choice of the initial conditions in the free expansion of the condensates open the possibility of planning specific scenarios that could be used, in principle, to obtain possible traces or signals caused by the quantum structure of space--time in low--energy earth--based experiments.

\begin{acknowledgments}
JIR acknowledges CONACyT grant No. 18176.
\end{acknowledgments}

\end{document}